\documentstyle[12pt]{article}
\date{}
\begin{document}

\title{ Quark Model Explanation of the $N^*\rightarrow
 N\eta$ Branching Ratios}

\author{L. Ya. Glozman$^{1,2,3}$ and D.O. Riska$^{4}$}
\maketitle

\centerline{\it $^1$Institute for Theoretical Physics,
University of Graz, 8010 Graz, Austria}

\centerline{\it $^2$Alma-Ata Power Engineering Institute,
480013 Alma-Ata, Kazakhstan}

\centerline{\it $^3$Research Institute for Theoretical Physics,
University of Helsinki,
00014 Finland}

\centerline{\it $^4$Department of Physics,
University of Helsinki,
00014 Finland}

\setcounter{page} {0}
\vspace{1.5cm}

\centerline{\bf Abstract}
\vspace{0.5cm}

The constituent quark model can explain the
strong selectivity of the $N\eta$ decay branching ratios
of the nucleon resonances if the fine structure interaction
between the constituent quarks is described in terms of
Goldstone boson exchange. This chiral quark model
predicts that the  resonances
$N(1535)$, $N(1710)$,
$\Lambda(1670)$, $\Sigma(1750)$, which have mixed flavor and spin
symmetry $[21]_{FS} [21]_F [21]_S$ wavefunctions in lowest
order, should have
large $N\eta$ branching ratios, while $N\eta$ decay of
the other
resonances that have
different flavor-spin symmetry should be strongly suppressed
in agreement with the experimental
branching ratios.\\
\\

Preprint HU--TFT--95--49

hep-ph

Submitted to Physics Letters B

\newpage
\normalsize

The relative magnitudes of the experimental $N\eta$ branching ratios
of the nucleon resonances reveal a number of outstanding and peculiar
features \cite{PDG,BATI}. The most notable is the size of the large
(30-50\%)
branching ratio for $N\eta$ decay of the lowest $L=1$,
$J^P={1\over 2}^-$ resonance
$N(1535)$
in comparison to the very small (1\%) $N\eta$ branching ratio
of the $N(1650)$, which is the following resonance with the
same quantum numbers. A similar feature is seen in the
spectrum of the $\Lambda$, where the large (15-35\%)
branching ratio
for $\Lambda\eta$ decay of the $L=1$, $J^P={1\over 2}^-$
$\Lambda(1670)$ resonance is followed
by a very small or vanishing $\Lambda\eta$ decay branching
ratio of the $\Lambda(1800)$. In contrast the $N\pi$
and $N\bar K $ branching ratios
of these resonances
show no comparable strong selectivity. \\

We here suggest a simple explanation for this strong
selectivity of the $N\eta$ decay
branching ratios, and of the similar selectivity of the corresponding
$\Lambda \eta$ and $\Sigma \eta$ branching ratios of the $\Lambda$ and
$\Sigma$ resonances that is based on the delicate interplay
between the fine structure and effective confining
interactions for the quarks, which obtains
when the former is described in terms of Goldstone boson (pseudoscalar meson)
exchange between the constituent quarks \cite{GLO1,GLO2,GLO3}.
Before proceeding we note however that some
of the notable differences
in the size of the $\eta$ decay branching ratios do have obvious
phase space explanations. Thus the very different $N\eta$ branching
ratios between the $N(1535)$, ${1\over 2}^-$ and the $N(1520)$,
${3\over 2}^-$ resonances can be explained by angular momentum
considerations alone: the $N\eta$ decay of the $N(1520)$ is
suppressed simply because the D-wave production mechanism that
is required for this decay is
suppressed in the neighborhood of the $N\eta$ threshold.
As the corresponding $N(1535) \rightarrow N\eta$ decay involves
an S-wave production mechanism it is not suppressed.
No such phase spase based explanation is however
possible for the suppression of
the $N\eta$ decay of the $N(1650)$, which has the same quantum numbers
as the $N(1535)$.\\

The conceptual framework for describing the fine structure interaction
between the constituent quarks in terms of the flavor dependent
pseudoscalar octet exchange interaction
between the constituent quarks \cite{GLO1,GLO2,GLO3}
is the two scale framework for QCD with 3 flavors \cite{MAG}.
According to this the appropriate degrees of freedom for describing
the dynamics between the chiral symmetry spontaneous breaking
scale $\Lambda_\chi^{-1}
\sim $ 0.2 -0.3 fm and the confinement scale $\Lambda_{QCD}^{-1}
\sim $ 1 fm are constituent quarks and the octet chiral fields.
It was shown in refs.\cite{GLO1,GLO2,GLO3} that a
phenomenological flavor-spin interaction
between the constituent quarks of the form associated with
the pseudoscalar octet exchange leads to a remarkably
good description of the known part
of the spectra of the nucleon, $\Delta$ and the strange hyperons.
The coupling between the constituent quark and the pseudoscalar mesons,
in this case governs the structure and content
of the quark sea has also been shown to resolve
several of the problems that in the naive quark model are
associated with the strangeness content
of the proton and the proton spin structure as measured in
deep inelastic lepton-proton scattering \cite{CHENG}.
The exceptional importance of the the octet of light pseudoscalar
mesons is their nature as the Goldstone
bosons of the approximate chiral symmetry of QCD.
We shall show that this model for the fine structure interaction
between the constituent quarks implies a clusterization into
quark-diquark configurations in baryon states with mixed
flavor symmetry with different quantum numbers and sizes of diquarks
in different baryons, and that this clusterization provides the
sought explanation of
the selectivity of the branching ratios for $\eta$ decay of the
baryon resonances.\\

Following \cite{GLO2,GLO3} we shall describe the effective confining
interaction between the constituent quarks by the harmonic oscillator
potential

$$V_{conf}=V_0+{1\over 6}m\omega^2\sum_{i<j}(\vec r_i-\vec
r_j),\eqno(1)$$

\noindent
where $m$ is the mass of the constituent quark  and
$\omega$ is the angular frequence of the oscillator
interaction. The unperturbed eigenvalues and eigenstates
of the 3-body Hamiltonian with the quark-quark interaction above are

$$E_0=3V_0+(N+3)\hbar\omega, \quad
\vert N (\lambda \mu) L [f]_X [f]_{FS} [f]_F [f]_S Y T>, \eqno(2)$$

\noindent
where $N$ is the number of excitation quanta in the state.\\

The spatial part of the three body wave function is determined by
the following quantum numbers:
$ N, (\lambda \mu), L,  [f]_X $.
Here we use the notations of the translationally invariant
shell model (TISM) \cite{Kurd}. The Elliott symbol $(\lambda \mu)$ determines
the SU(3) harmonic oscillator multiplet
 and $L$ is the total orbital angular momentum.
The spatial permutational symmetry of the state
is indicated by the Young pattern $[f]_X$, where f is
a sequence of integers that indicate the number
of boxes in the successive rows of the corresponding Young patterns.
The $[f]_F$ , $[f]_S$ and $[f]_{FS}$
 Young patterns denote the permutational flavor,
spin and combined flavor-spin
symmetries respectively. Note that the totally antisymmetric color
state $[111]_C$, which is common to all the states, has been
suppressed in (2). By the Pauli Principle $[f]_X=[f]_{FS}$.
All the required wave functions are given explicitly in \cite{GLOK}.

The explicit zero order wave functions (2)
for the nucleon , the $N(1535)$ and the $N(1650)$, which have the
symmetry structures
$\vert 0 (00) 0 [3]_X [3]_{FS} [21]_F [21]_S 1 {1\over2}>$,

\noindent
$\vert 1 (10) 1 [21]_X [21]_{FS} [21]_F [21]_S 1 {1\over2}>$
and
$\vert 1 (10) 1 [21]_X [21]_{FS} [21]_F [3]_S 1 {1\over2}>$
respectively are

$$\Psi_N ={1 \over \sqrt{2}}
 \varphi_{000}(\vec {r}_{12})\varphi_{000}(\vec {\rho})$$

$$\times \left[ \vert S_{12}T_{12}=00, S_3T_3=\frac {1}{2}\frac {1}{2}
:ST=\frac {1}{2}\frac {1}{2}>
+ \vert S_{12}T_{12}=11, S_3T_3=\frac {1}{2}\frac {1}{2}
:ST=\frac {1}{2}\frac {1}{2}>\right], \eqno(3a)$$

$$\Psi_{N(1535)} =\sum_{M,M_S} \frac {1}{2}
<1M \frac{1}{2}M_S\vert \frac{1}{2} M_J>
\left\{ \varphi_{000}(\vec {r}_{12})\varphi_{11M}(\vec {\rho}) \right.$$

$$\times \left[ \vert S_{12}T_{12}=00, S_3T_3=\frac {1}{2}\frac {1}{2}
:ST=\frac {1}{2}\frac {1}{2}>
- \vert S_{12}T_{12}=11, S_3T_3=\frac {1}{2}\frac {1}{2}
:ST=\frac {1}{2}\frac {1}{2}>\right]$$

$$ + \varphi_{11M}(\vec {r}_{12})\varphi_{000}(\vec {\rho})$$

$$\left. \times \left[ \vert S_{12}T_{12}=01, S_3T_3=\frac {1}{2}\frac {1}{2}
:ST=\frac {1}{2}\frac {1}{2}>
+ \vert S_{12}T_{12}=10, S_3T_3=\frac {1}{2}\frac {1}{2}
:ST=\frac {1}{2}\frac {1}{2}>\right] \right\}, \eqno(3b)$$

$$\Psi_{N(1650)} = \sum_{M,M_S} \frac {1}{\sqrt{2}}
<1M \frac{3}{2}M_S\vert \frac{1}{2} M_J>$$

$$\times \left\{ \varphi_{000}(\vec {r}_{12})\varphi_{11M}(\vec {\rho})
\vert S_{12}T_{12}=11, S_3T_3=\frac {1}{2}\frac {1}{2}
:ST=\frac {3}{2}\frac {1}{2}>\right.$$

$$\left. + \varphi_{11M}(\vec {r}_{12})\varphi_{000}(\vec {\rho})
\vert S_{12}T_{12}=10, S_3T_3=\frac {1}{2}\frac {1}{2}
:ST=\frac {3}{2}\frac {1}{2}>\right\}.\eqno(3c)$$

\noindent
Here $S_{12}$,$S_3$ and $S$ denote spin of the particle pair (1 2),
the particle 3 and the total spin respectively.
$T_{12}$, $T_3$ and $T$ denote the corresponding isospins.
The Jacobi coordinates
$\vec {r}_{12}$ and $\vec {\rho}$ are defined as:

$$\vec {r}_{12} = \frac {\vec {r}_1 - \vec{r}_2} {\sqrt {2}}, \quad
\vec {\rho} = \frac {\vec {r}_1 + \vec{r}_2 - 2\vec{r}_3} {\sqrt {6}}.
\eqno(4)$$

\noindent
and finally $\varphi_{nlm}$ are the harmonic oscillator functions
with $n$ excited quanta. \\

In an effective chiral symmetric description of baryon
structure based on the constituent quark model
the coupling of the quarks and the pseudoscalar Goldstone
bosons will (in the $SU(3)_F$ symmetric approximation) have
the form $ig\bar\psi\gamma_5\vec\lambda^F
\cdot \vec\phi\psi$, where $\psi$ is the fermion constituent quark
field operator, $\vec\phi$ is the octet boson field
operator,  g is a flavor independent
coupling constant, and ${\vec{\lambda}}^F$
are the $SU(3)$ flavor Gell-Mann matrices. To lowest order in the
nonrelativistic reduction for a constituent quark spinor
this coupling will (neglecting the recoil correction)
give rise to the following vertex operators:

$$ \hat {O}_{\pi q} \simeq {g\over 2m} \vec {\tau}
\vec {\sigma} \cdot \vec {q}
e^{i \vec{q}  \vec{r}}, \quad \hat {O}_{\eta q} \simeq {g\over 2m}
\lambda^8 \vec {\sigma}\cdot \vec {q}
e^{i \vec {q}  \vec{r}}, \eqno(5)$$

\noindent
where $\vec {\tau}$ are isospin Pauli matrices and $\vec{q}$ is meson momentum.
As $\lambda^8$ is diagonal $\eta$-absorption or emission does not change
the quark flavor.\\

With the operators (5) and the basis functions (3) the
transition matrix elements $N(1650) \rightarrow
N\pi$, $N(1650) \rightarrow N\eta$, $N(1535) \rightarrow N\pi$,
$N(1535) \rightarrow N\eta$ have the form

$$T=< N \vert \sum _{i=1}^3 \hat {O}(i) \vert N^*>. \eqno (6)$$

\noindent
As all of these are of the same order and large it is impossible
to explain the very strong suppression of
the $N(1650) \rightarrow N\eta$ decay at
this level without taking into account the effect of the
fine structure interaction between the quarks.\\

In the $SU(3)_F$ symmetric approximation
the key component of the effective chiral boson exchange
interaction between the constituent quarks is

$$H_\chi\sim -\sum_{i<j}V(\vec r_{ij})
\vec \lambda^F_i \cdot \vec \lambda^F_j\,
\vec
\sigma_i \cdot \vec \sigma_j.\eqno(7)$$
In the description of the baryon spectrum in ref.
\cite{GLO2,GLO3}
the matrix elements of the interaction potential $V(r_{ij})$
in the lowest harmonic oscillator states were extracted from
the 4 lowest splittings in the nucleon and $\Delta$ spectrum,
while the detailed radial behavior of it was left unspecified.
At large interquark separation the potential $V(r_{ij})$
will asymptotically behave as a negative Yukawa function with
a falloff that is determined by the appropriate meson ($\pi,K,\eta$)
mass. At short range $r_{ij}<0.6-0.7$fm the potential will
on the other hand be positive and large in magnitude, reminiscent
of a smeared version
of the $\delta$-function in the pion exchange potential. The
average strength of this positive short range potential is
given by the S-state oscillator matrix element $P_{00}=
<000\vert V(r)\vert 000>$, which is determined by the
$N\Delta$ splitting $10P_{00}$ to be 29.3 MeV \cite {GLO2,GLO3}.\\

The flavor-spin matrix elements of the chiral field interaction
in light quark pair states with spin $S_{ij}$ and isospin $T_{ij}$
are

$$<[f_{ij}]_F\times [f_{ij}]_S : [f_{ij}]_{FS}
{}~| -V(r_{ij})\vec \lambda^F_i \cdot \vec \lambda_j^F
\vec\sigma_i \cdot \vec \sigma_j
{}~|~[f_{ij}]_F \times [f_{ij}]_S : [f_{ij}]_{FS}>$$
$$=\left\{\begin{array}{rr} -{4\over 3}V(r_{ij})& [2]_F,[2]_S:[2]_{FS}
{}~(S_{ij}=1, T_{ij}=1) \\
-8V(r_{ij}) & [11]_F,[11]_S:[2]_{FS} ~(S_{ij}=0, T_{ij}=0) \\
4V(r_{ij}) & [2]_F,[11]_S:[11]_{FS} ~(S_{ij}=0, T_{ij}=1)\\ {8\over
3}V(r_{ij}) & [11]_F,[2]_S:[11]_{FS} ~(S_{ij}=1, T_{ij}=0)
\end{array}\right.\eqno(8)$$

\noindent
The one-to-one correspondence between the flavor permutational
symmetry $[f_{ij}]_F$ and the total isospin $T_{ij}$ is valid only
in the u and d quark subspace.\\

It is instructive to compare the matrix elements of the effective
confining interaction (1) to those of the fine structure interaction
in the harmonic oscillator ground state. The former is

$$ <\varphi_{000}(\vec {r}_{ij}) \vert V_{conf}(\vec {r}_{ij})
\vert \varphi_{000}(\vec {r}_{ij})> = V_0 + \frac {3}{4} \hbar
\omega =
70.4 \,\,{\rm MeV},\eqno(9)$$
where $\hbar \omega =$ 157.4 MeV and $V_0=$--47.7 MeV
\cite{GLO2,GLO3}.\\

\noindent
The corresponding matrix element of the chiral field interaction
$(P_{00})$ (29.3 MeV) in the $S_{ij}=T_{ij}=0$ state is multiplied
by the factor $ -8$ , so that in fact

$$  <\varphi_{000}(\vec {r}_{ij}), S_{ij}T_{ij}=00  \vert
 -V(r_{ij})\vec \lambda^F_i \cdot \vec \lambda_j^F
\vec\sigma_i \cdot \vec \sigma_j
\vert \varphi_{000}(\vec {r}_{ij}), S_{ij}T_{ij}=00 >$$

$$  = - 8P_{00} = -234.4 \,\,{\rm MeV}. \eqno(10)$$

\noindent
As the confining interaction (1) is monotonically rising function and
$V(r_{ij})$ in (7) changes sign in the vicinity of $r_{ij} \sim
0.6-0.7$ fm this comparison
of matrix elements indicated
that at short distances ($r\le 0.6-0.7$ fm)
the confining interaction is weak in comparison to the
chiral field interaction.
As one half of the weight
of the zero order nucleon state (3a) is formed of quark
pair states with $S_{12}=T_{12}=0$ the consequence
of the strong dominance of the attractive chiral field
interaction at short distances
in the $S_{12}=T_{12}=0$ pair in the nucleon ground state
the quarks will clusterize
into a quark-diquark configuration,
the latter one having $S_{12}=T_{12}=0$ and being very compact.
The chiral field interaction is attractive, but weaker by a
factor 6
in the $S_{12}=T_{12}=1$ pair state, which consequently will
be strongly suppressed. A similar quark-diquark
clustering will occur in the
ground states of the other octet baryons, where in general the
compact diquark will be characterized by antisymmetric spin
and flavor wavefunctions.
No such clusterization
will on the other hand occur in the decuplet baryons, as these
have no quark pairs with antisymmetric flavor symmetry.\\

Consider now the zero order $N(1650)$ wave function in (3c).
The quark pair components
in this state have $S_{12}=T_{12}=1$ and $S_{12}=1, T_{12}=0$.
In the former there is a rather soft attractive interaction and in the
latter there is a much larger repulsive one (the
matrix elements of the potential $V(r)$ are of similar sign
and magnitude in $S$ and in $P$ states). The net effect
of the "pull" in the former and the "push" in the latter will
again lead to a clusterization into a quark-diquark
configuration at short distances, where the confining
interaction is weak, but in this case the diquark will have
$S_{12}= T_{12}=1$. The size of this diquark will be considerably
larger than that of the $S_{12}=T_{12}=0$
because of the weaker net attractive
clusterforming interaction in the $S_{12}=T_{12}=1$ channel.\\

This clusterization into quark-diquark configurations with
different quantum numbers in the baryon resonances with
mixed flavor symmetry $[21]_F$ provides the explanation
for the suppression of the $N\eta$ decay branch of the
$N(1650)$ and for the large $N\eta$ branch
of the $N(1535)$. Consider first the case of the $N(1650)$.
As in $\eta$ decay the isospin of the involved quark pair
state is unchanged it cannot proceed through the
$S_{12}=1,T_{12}=1\rightarrow S_{12}=0,T_{12}=0$
transition that would connect the diquark states in the
$N(1650)$ and the nucleon.
The corresponding pion decay, in which isospin flip
is possible, can on the other hand connect these pair
states. The reason for why the branching ratio for the
transition $N(1650)\rightarrow
N\eta$ does not vanish entirely
is that the suppression at short range of the
the pair states that do not form diquarks in the nucleon
and the $N(1650)$ is not complete.
The reason for the large $\eta$ decay branch of the $N(1535)$
is that its zero order wavefunction (3b) has components with the
same spin- and flavor structure as the nucleon (i.e. there is
also clusterization with $S_{12}=T_{12}=0$ diquark quantum numbers), and
hence the quark-diquark clusterization has no consequence
for the $N(1535)\rightarrow N\eta$ decay branch.\\

This argument generalizes to the predictions that:
(1) all baryon resonances with
$[21]_{FS}[21]_F[21]_S$ flavor-spin symmetry
zero order wavefunctions and smallest possible
angular momentum should have large $\eta$-decay branching
ratios whereas
(2) the baryon resonances that have $[21]_{FS}[21]_F[3]_S$ symmetry
zero
order wave functions should have strongly suppressed
$\eta$-decay branching ratios.\\

The confirmed nucleon and $\Lambda$ states that are characterized
by the $[21]_{FS}[21]_F[21]_S$ flavor-spin symmetry are
\cite {GLO2,GLO3}

$$\frac {1}{2}^-, N(1535); \frac {3}{2}^-, N(1520);
\frac {1}{2}^+, N(1710);$$
$$\frac {1}{2}^-, \Lambda(1670); \frac {3}{2}^-,
\Lambda(1690).$$

\noindent
The $J=\frac {1}{2}$ states of these should
therefore have large $N\eta$ branching ratios.
This prediction is in excellent agreement with the
corresponding empirical branching ratios, all of which are
large \cite{PDG,BATI}:

$$\frac {\Gamma_{N(1535) \rightarrow N\eta}}{\Gamma_{total}} =
30-50\%,\quad
\frac {\Gamma_{N(1710) \rightarrow N\eta}}{\Gamma_{total}} =
20-40\%,$$
$$\frac {\Gamma_{\Lambda(1670) \rightarrow \Lambda\eta}}
{\Gamma_{total}} = 15-35\% .$$

\noindent
The confirmed baryon resonances that have $[21]_{FS}[21]_F[3]_S$
flavor
symmetry zero order wavefunctions
are

$${\frac {1}{2}}^-, N(1650); {\frac {3}{2}}^-, N(1700);
{\frac {5}{2}}^-, N(1675);$$

$${\frac {1}{2}}^-, \Lambda(1800); {\frac {5}{2}}^-, \Lambda(1830).$$

The empirical $\eta$-decay branching ratios for these resonances
do not exceed 1\%, which agrees with the general prediction above.
\\

Among the $\Sigma$ resonances only the
the ${\frac{1}{2}}^-, \Sigma(1750)$ has a large branching ratio
(15-35\%) for $\Sigma \eta$ decay. This then implies
the flavor symmetry of the zero order wave function for this state
should be $[21]_{FS} [21]_F [21]_S$ when the argument above
is turned around. Moreover the state with this flavor spin
symmetry should on the basis of the spectral predictions in
ref. \cite{GLO3} be the lowest $\frac{1}{2}^-$ state in the
$\Sigma$ spectrum. As a corollary the lower lying
tentative (2 star) negative parity resonances
$\Sigma(1580)$ and $\Sigma(1620)$ are likely to be spurious.
This conclusion is of course hinged on the large
empirical branching ratio for $\Sigma(1750) \rightarrow
\Sigma \eta$.\\

To summarize we have presented a qualitative explanation
of strong selectivity displayed by the $\eta$-decay branching
ratios of different  baryon resonances.
The explanation is closely tied to the description of the
structure of the baryons in terms of a fine structure interaction
that is mediated by the octet of light pseudoscalar mesons,
which play a special role as the Goldstone bosons of the
approximate chiral symmetry of QCD. The key point of this
explanation for the selectivity of the branching ratios for
$\eta$ decay is the relative smallness of the confining
interaction at short range in the 3-quark wavefunctions
of the octet baryons, and consequent clusterization
into quark-diquark configurations driven by the chiral
boson exchange interaction. Moreover we have proposed a
definite symmetry assignment for the $\Sigma(1750)$ and
and an argument for the spuriosity of the
2-star $\Sigma(1580)$ and $\Sigma(1620)$ resonances.
We note finally that it is the suppressed $N\eta$ decay
of the $N(1650)$ as compared to the large $N\eta$ branching
fractions of the nearby $N(1535)$ and $N(1710)$ resonances
that presents the main difficulty for finding
an explanation for the selectivity
of the $\eta$ decay branching ratios in terms of a phenomenological
baryon field theory, which does not take into account the
internal structure of the baryon resonances.


\

\begin{thebibliography}{999}
\bibitem {PDG} Particle Data Group, Phys. Rev. {\bf D45} (1992) II,
Phys. Rev. {\bf D50} (1994) 1173
\bibitem{BATI} M. Batinic et al., $\pi N\rightarrow \eta N$ and
$\eta N\rightarrow \eta N$ Partial Wave T-matrices in a
Coupled, Three Channel Model, UCLA-10-P25-230 (1995)
\bibitem{GLO1} L. Ya. Glozman and D. O. Riska,
The Baryon Spectrum and Chiral Dynamics, Preprint HU-TFT-94-47,
LANL hep-ph 9411279 (1994)
\bibitem{GLO2} L. Ya. Glozman and D. O. Riska,
Systematics of the Light and Strange Baryons and the
Symmetries of QCD, Preprint HU-TFT-94-48,
LANL hep-ph 9412231 (1994)
\bibitem{GLO3} L. Ya. Glozman and D. O. Riska,
The Spectrum of the Nucleons and the Strange Hyperons
and Chiral Dynamics, Preprint DOE/ER/40561-187-INT95-16-01,
LANL hep-ph 9505422, Physics Reports (in press)
\bibitem{MAG} A. Manohar and H. Georgi, Nucl. Phys. {\bf B234}
 (1984) 189
\bibitem{CHENG} T. P. Cheng and L.-F. Li, Phys. Rev. Lett.
{\bf 74} (1995) 2872
\bibitem{Kurd} I. V. Kurdyumov, Yu. F. Smirnov, K. V. Shitikova
and K. El. Samarai, Nucl. Phys. {\bf A145} (1970) 593
\bibitem{GLOK} L. Ya. Glozman and E. I. Kuchina,
Phys. Rev. {\bf C49} (1994) 1149
\end{thebibliography}
\end{document}